\documentclass{svmult}

\usepackage{mathptmx}       
\usepackage{helvet}         
\usepackage{courier}        
\usepackage{type1cm}        
\usepackage{overpic}
\usepackage{makeidx}         
\usepackage{graphicx}
\usepackage[titletoc,title]{appendix}        
\usepackage[dvipsnames]{xcolor}							 
\usepackage{multicol}        
\usepackage[bottom]{footmisc}
\usepackage{amsmath,amssymb,epsfig}
\usepackage{refcount}

\makeindex             

\begin{document}

\title*{On homotopy continuation based singularity distance computations for 3-RPR manipulators}
\titlerunning{On homotopy continuation based singularity distance computations} 
\author{Aditya Kapilavai \and Georg Nawratil}
\authorrunning{A.\ Kapilavai and G.\ Nawratil}
\institute{Institute of Discrete Mathematics and Geometry, TU Wien, Austria \\ \email{\{akapilavai, nawratil\}@geometrie.tuwien.ac.at}} 	
	
\maketitle

\abstract{It is known that parallel manipulators suffer from singular configurations. 
Evaluating the distance between a given configuration to the closest singular one is of interest for 
industrial applications (e.g.\ singularity-free path planning). 
For parallel manipulators of Stewart-Gough type, geometric meaningful distance measures are known, which are used for the computation of the singularity distance as the global minimizer of an optimization problem. 
In the case of hexapods and linear pentapods the critical points of the corresponding polynomial Lagrange function cannot be found by the Gr\"obner basis method 
due to the degree and number of unknowns. 
But this polynomial system of equations can be solved by software tools of numerical algebraic geometry relying on homotopy continuation. 
To gain experiences for the treatment of the mentioned spatial manipulators, 
this paper attempts to find minimal multi-homogeneous B\'{e}zout numbers 
for the homotopy continuation based singularity distance computation  with respect to various algebraic 
motion representations of planar Euclidean/equiform kinematics. The homogenous and non-homogenous representations under study are 
compared and discussed based on the 3-RPR manipulator.}
\keywords{3-RPR manipulator, singularity distance, homotopy continuation, B\'{e}zout number}

\section{Introduction}
\label{sec:introduction}
\begin{figure}[t]
\begin{center}
\begin{overpic}[width = 84mm]{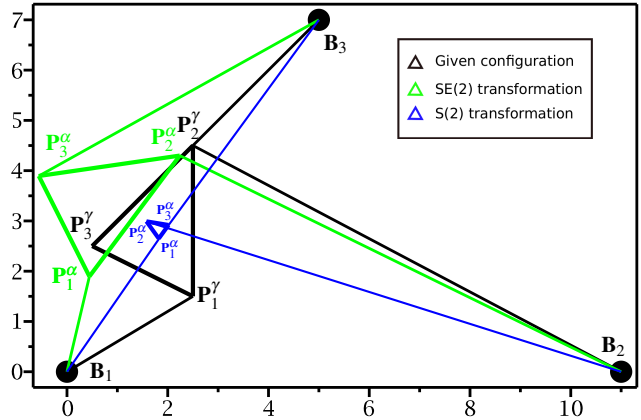}
\begin{small}‎
\put(13,8){$\mathbf{B}_{1}$}‎
\put(30,20){$\mathbf{P}^\gamma_{1}$}‎
\put(94,12){$\mathbf{B}_{2}$}‎
\put(26.8,47){$\mathbf{P}^\gamma_{2}$}‎
\put(9.9,31.1){$\mathbf{P}^\gamma_{3}$}‎
\put(50,60){$\mathbf{B}_{3}$}‎
\put(7,23){\color{green}$\mathbf{P}^\alpha_{1}$}
\put(22.7,45){\color{green}$\mathbf{P}^\alpha_{2}$}‎
\put(6,44.1){\color{green}$\mathbf{P}^\alpha_{3}$}‎
\put(7,23){\color{green}$\mathbf{P}^\alpha_{1}$}
\put(24.3,28){\tiny\color{blue}$\mathbf{P}^\alpha_{1}$}
\put(23.6,34){\tiny\color{blue}$\mathbf{P}^\alpha_{3}$}
\put(19.2,30.3){\tiny\color{blue}$\mathbf{P}^\alpha_{2}$}
\end{small}‎
\end{overpic}
\caption{
3-RPR manipulator: The results obtained in this paper are based on the illustrated example taken from \cite[Section 3]{Nawratil_2019} for $\phi=\frac{\pi}{2}$. 
The given configuration is illustrated in black and the closest singular one under Euclidean/equiform transformations of the platform is displayed in green/blue.
}
\label{fig1}
\end{center}
\end{figure}

A 3-RPR manipulator (cf.\ Fig.~\ref{fig1}) is a three degree-of-freedom (dof) planar parallel manipulator (two translational and one rotational dof) where each leg is composed of a revolute joint (R)\footnote{We consider R-joints as points and refer them as base/platform anchor points.}, a prismatic joint (P) and a further R-joint. The manipulator is actuated by changing the three lengths of the prismatic joints. 

According to~\cite{Nawratil_2019}, the distance between two poses of the moving platform 
can be computed as:
\begin{equation}
d_3 := \sqrt{\frac{1}{3}\sum_{i=1}^{3} \langle \mathbf{P}_{i}^\alpha-\mathbf{P}_{i}^\gamma,  \mathbf{P}_{i}^\alpha-\mathbf{P}_{i}^\gamma  \rangle} 
\label{eq:distance}
\end{equation}
where $\mathbf{P}_{i}^\gamma = (x_i^\gamma, y_i^\gamma)^T$ (resp.\ $\mathbf{P}_{i}^\alpha=({x_i}^\alpha, {y_i}^\alpha)^T$) 
denotes the $i$th platform anchor point in the given (resp.\  $\alpha$-transformed) manipulator configuration for $i=1,2,3$. 
Both coordinate vectors $\mathbf{P}_{i}^\gamma$ and $\mathbf{P}_{i}^\alpha$ 
are computed with respect to the fixed frame and $\langle \,,\,\rangle$ denotes the standard scalar product. The coordinate vectors 
of the platform anchor points with respect to the moving frame 
are given by  $\mathbf{P}_{i} = (x_i, y_i)^T$. 
Moreover it was mentioned in \cite{Nawratil_2019} that $\alpha$ has not to be restricted to the planar Euclidean motion group SE(2), but it can also be an element of the 
planar equiform motion group S(2) or an affine motion.

It is well known that the $\alpha$-transformed configuration is singular if and only if the carrier lines of three legs intersect in a common point or are parallel. 
This line-geometric characterization is equivalent to the algebraic condition $V_3=0$ with 
 \begin{equation}
\mathrm{V_3} = \det\begin{pmatrix}
\mathbf{P}_{1}^\alpha-\mathbf{B}_{1} & \phantom{-}\mathbf{P}_{2}^\alpha-\mathbf{B}_2 & \phantom{-}\mathbf{P}_{3}^\alpha-\mathbf{B}_3 \\ 
\det\left(\mathbf{B}_{1},\mathbf{P}_{1}^\alpha-\mathbf{B}_{1}\right)& \phantom{-}\det\left(\mathbf{B}_{2}, \mathbf{P}_{2}^\alpha-\mathbf{B}_{2}\right)& \phantom{-}\det\left(\mathbf{B}_{3}, \mathbf{P}_{3}^\alpha                                                                    - \mathbf{B}_{3}\right) \\
  
\end{pmatrix}
 \label{eq:s}
\end{equation}
where $\mathbf{B}_i$ denotes the coordinate vector of the $i$th base anchor point with respect to the fixed system. For determining the singularity distance one has to find the 
transformation $\alpha$ that minimizes Eq.\ (\ref{eq:distance}) under the side condition that $V_3=0$ holds.
For $\alpha\in$ SE(2) we denote the singularity distance by   $s_3$ and for $\alpha\in$ S(2) by $e_3$, 
respectively. 
For the determination of $s_3$ and $e_3$ we compute the critical points of the corresponding polynomial 
Lagrange function $L$, i.e.\ the zero set of the polynomial system of equations arising from the partial derivatives of $L$. 
Finally, one only has to pick out the solution which yields the global \smallskip minimum.

Commonly, Newton's method is simple and quick to solve a polynomial system. However, in general it only converges when a good initial guess is used, and even then this only yields one solution. In contrast, the homotopy continuation method allows an efficient and reliable computation of all solutions to polynomial systems. The newest open-source 
numerical continuation software packages are HomotopyContinuation.jl~\cite{10.1007/978-3-319-96418-8_54} and Bertini 1.6v~\cite{BHSW06}. These packages and other available software and their performances are compared in \cite{10.1007/978-3-319-96418-8_54}. We have chosen Bertini 1.6v for computing the singularity distance of parallel manipulators due to 
(a) its exceptional features mentioned in~\cite [p.~156]{10.1007/978-1-4471-4510-3_5} which will be useful for our future research and (b) the longevity of the Bertini software.
 
Under the default settings, declaration of all unknown variables $n$ of the polynomial system into a single {\it variable group} causes Bertini to form a {\it total-degree homotopy}, 
which in general does not result in a minimal B\'{e}zout number.
The partition of the variables into multiple groups lead to a {\it multi-homogeneous homotopy}. The grouping of the $n$ unknown variables affects the resulting minimal multi-homogeneous B\'{e}zout number\footnote{The B\'{e}zout number of a system of multi-homogeneous polynomial equations is the largest number of non-singular solutions such a system can have, and it is also the number of solution paths used to compute all geometrically isolated solutions of the system using multi-homogeneous polynomial continuation~\cite{Wampler1992}.} ($B_{min}$). 
As Bertini does not group the variables automatically, the user is responsible for choosing the variable groups in a way that a minimal multi-homogeneous B\'{e}zout number ($B_{min}$)
is obtained, which improves the computational costs/time. 
For this purpose, attempts have been made on search algorithms (e.g.\ \cite{Wampler1992, Li2001}) to find the optimal partition of variables groups, but in \cite [p.~72]{bates2013numerically} it is stated that \textit{``there does not yet exist a truly efficient algorithm for finding optimal groupings,  
and the combinatorics are such that an exhaustive examination of all possible groupings becomes impractical as the number of variables grows much larger than 10''}. 
Investigating the optimal variable group search algorithms is out of the scope of the paper, but as mentioned in~\cite [p.~73]{bates2013numerically}, \textit{``with a little practice, a user can identify and check the most promising candidates''} resulting in  $B_{min}$. 

According to \cite [p.~96]{bates2013numerically} the \textit{``use of isotropic coordinates is advantageous when applying multi-homogeneous homotopy, since it converts many of the quadratic 
expressions that appear in planar kinematics into bilinear expressions''} but no attempts have been made so far to compare this approach with other algebraic motion representations. 
Hence, we determine $B_{min}$ for the singularity distance computation of 3-RPR manipulators based on homotopy continuation with respect to various algebraic 
motion representations of SE(2) and S(2). In addition this study is useful to gain experience for the spatial case.

The rest of the paper is organized as follows: We divide the investigated algebraic motion representations into two classes, namely the non-homogenous ones (Sec.\ \ref{NHR})
and the homogenous ones (Sec.\ \ref{HR}). In both of these two sections we give a brief introduction of the representations followed by the discussion of the results. 
Finally, Section \ref{conc} concludes the paper.


\section{Non-Homogeneous Representations}
\label{NHR} 
In this section, we present three non-homogeneous algebraic representations for each of the motion groups SE(2) and S(2), \medskip respectively.

\noindent
\textbf{Point Based Representation (PBR):}
We call the approach given in~\cite{Nawratil_2019} {\it point based representation} as the transformation is formulated in terms of the 
first and second platform anchor points. For the third platform point the transformation reads as

\begin{equation}
\mathbf{P}_{3}^\alpha=  \tfrac{1}{\sqrt{(x_2-x_1)^{2}+(y_2-y_1)^{2}}} \begin{pmatrix}
x_2^\alpha-x_1^\alpha & \phantom{-}y_1^\alpha-y_2^\alpha\\ 
y_2^\alpha-y_1^\alpha &  \phantom{-}x_2^\alpha-x_1^\alpha
\end{pmatrix}   
\begin{pmatrix}
x_3-x_1 \\
y_3-y_1
\end{pmatrix} 
    +\mathbf{P}_{1}^\alpha.
\end{equation}
Then the Lagrange function $L$ for the computation of $e_3$ can be written as 
\begin{equation}
{L}:\,\, d_3^2-\lambda V_3=0.
\label{s3} 
\end{equation}
The additional constraint $M=0$ with $M:=\overline{\mathbf{P}_{1}^\alpha\mathbf{P}_{2}^\alpha}^2-\overline{\mathbf{P}_{1}\mathbf{P}_{2}}^2$ 
results in the computation of $s_3$ from the Lagrange function
\begin{equation}\label{points5b}
{L}:\,\, d_3^2-\lambda V_3-\mu M=0.
\end{equation}

\noindent
\textbf {Planar Euler-Rodrigues Representation (PERR):}
\label{sssec:num2}
Using this representation the transformation $\alpha:\,\, \mathbb{R}^2 \to \mathbb{R}^2$ can be written as
\begin{equation}
\alpha:\,\, \mathbf{P}_{i} \mapsto \mathbf{P}_{i}^\alpha:= \begin{pmatrix}
a_1^2-a_2^2 & \phantom{-}-2a_1a_2 \\ 
2a_1a_2 &  \phantom{--}a_1^2-a_2^2
\end{pmatrix} \mathbf{P}_{i}+
\begin{pmatrix}
a_3 \\ a_4
\end{pmatrix} \quad \text{with} \quad a_1,\ldots, a_4\in\mathbb{R}.
\label{eq:113}
\end{equation}
The computation of $e_3$ and $s_3$ is based on the same Lagrange function as for the PBR with the sole difference that $M$ is given 
by \medskip $M:=a_1^2+a_2^2-1$. 

\noindent
\textbf {Isotropic Coordinates Representation (ICR):} 
Using this approach the point $\mathbf{P}_i$ is represented by the pair $(z_i,\overline z_i)$ of conjugate complex numbers with $z_i=x_i+iy_i$. This  
isotropic coordinates are transformed by $\alpha:\,\, \mathbb{C}^2 \to \mathbb{C}^2$ as follows (e.g.\ \cite{nawratil2018parallel}):
\begin{equation}
\alpha:\,\, ({z_i},\overline z_i) \mapsto (z_i^\alpha,\tilde z_i^\alpha):=(\kappa {z_i}+ {\tau}, \tilde\kappa {\overline z_i}+ {\tilde\tau}) \quad \text{with} 
\quad \kappa,\tau, \tilde\kappa,\tilde\tau \in\mathbb{C}.
\end{equation}
Note that $(z_i^\alpha,\tilde z_i^\alpha)$ is a real point if and only if $\tilde z_i^\alpha=\overline z_i^\alpha$ holds; i.e.\ 
$\tilde\kappa=\overline\kappa$ and $\tilde\tau=\overline\tau$. 
Again the computation of $e_3$ and $s_3$ is based on the same Lagrange function as for the PBR with the sole difference that $M$ is given 
by \medskip $M:=\kappa\tilde{\kappa}-1$.

In summary the transformation $\alpha$ is given by the non-homogenous 4-tuple 
$(x_1^\alpha,y_1^\alpha,x_2^\alpha,y_2^\alpha)$ in PBR, $(a_1,a_2,a_3,a_4)$ in PERR and $(\kappa,\tau, \tilde\kappa,\tilde\tau)$ in ICR, respectively. 

\begin{table}[b]
  \centering
  \caption{Comparison of B\'{e}zout numbers for all studied representations of SE(2)}
  \begin{tabular}{|c||c|c|c|c|c|c|} \hline
  Method &  Best groupings  & $B_{min}$  & $T_{avg}$  & Worst groupings &  $B_{max}$ & DOL  \\
    \hline\hline
    PBR & $\{{(x_1^\alpha, y_1^\alpha, x_2^\alpha, y_2^\alpha)}, (\lambda, \mu)\}$ &  144  & 1\,800 & $\{{(x_1^\alpha), (x_2^\alpha, \mu),(y_1^\alpha, \lambda),(y_2^\alpha)}\}$  & 8\,448 & 4\\
     &  &    &  & $\{(x_1^\alpha, \mu), (y_2^\alpha, \lambda),(x_2^\alpha),(y_1^\alpha)\}$  & 8\,448 & \\

    PERR & $\{(a_1, a_2),(a_3, a_4),(\lambda,\mu)\}$  & 360 & 45\,920 & $\{(a_1,\lambda, a_3),(a_2, \mu, a_4)\}$ & 50\,992 & 6 \\
		     & $\{(a_1, a_2,a_4), (a_3),(\lambda,\mu)\}$  & 360 & 35\,640 &  $\{(a_1,\mu, a_4),(a_2, \lambda, a_3)\}$ & 50\,992 &  \\
				 
    ICR & $\{(\kappa), (\tilde{\kappa}), (\lambda), (\mu), (\tau,\tilde{\tau})\}$  &  136 & 249\,880& $\{(\kappa, \tilde{\tau}),(\tilde{\kappa}, \mu, \tau, \lambda)\}$   & 2\,187 & 4 \\
		
		    &   &   & &  $\{(\kappa, \mu, \lambda,\tilde{\tau}),(\tilde{\kappa}, \tau)\}$   & 2\,187 &  \\

   \hline
   BGR  &  $\{{(e_0, e_3,t_1, t_2)_h, (\lambda)\}}$       & 300  & 122\,180  &  --  & -- & $\tfrac{5}{4}$\\
    DCKR  &  $\{(\theta, \tilde{\theta}, \sigma, \tilde{\sigma})_h, (\lambda)\}$        & 300  & 115\,283 &  --  & -- & $\tfrac{5}{4}$ \\
    
    \hline
  \end{tabular}
  \label{tab:5}
\end{table}

\subsection{Results}
\label{RNHR}
The results are based on 
the example illustrated in Fig.~\ref{fig1}. 
The system of ${n}$ partial derivatives ${L_j}$~$(j=1,\dots, n)$ of ${L}$ results in all cases in a 
non-homogenous system of  equations. 
The number $n$ of unknown variables in this polynomial system is $5$ for $\alpha\in$ S(2) and $6$ for $\alpha\in$ SE(2). 
The total number of all possible groupings of the $n$ variables is given by the so-called {\it Bell number}\footnote{In case if $n > 6$ 
the Bell number ${B(n)}$ increases very quickly (see~\cite[p.~72, Table 5.1]{bates2013numerically}).}  ${B(n)}$ with $B(5) = 52$ and $B(6)= 203$.
For every representation all the possible groupings are tested by using default settings in Bertini 1.6v with a multi-homogeneous homotopy continuation method. 
We used Linux OS (Ubuntu 18.04) with a 1.80GHz Intel i5-6260U CPU. 
Comparison of  B\'{e}zout numbers and degree of the Lagrange function (DOL) 
for all the presented representations are summarized in the Tables \ref{tab:5} and \ref{tab:6}, where 
$B_{min}$ and $B_{max}$ indicates the B\'{e}zout number for best and worst groupings of unknown variables. 
We are also interested in average total tracking time $T_{avg}$ in Milliseconds for $B_{min}$. 
We took $T_{avg}$ because there will be variations in total tracking time for each run, depending upon the random seed chosen by Bertini.
 \begin{table}[t]
  \centering
  \caption{Comparison of B\'{e}zout numbers for all studied representations of S(2)}
  \begin{tabular}{|c||c|c|c|c|c|c|} \hline
  Method & Best groupings & $B_{min}$ & $T_{avg}$ & Worst groupings & $B_{max}$ & DOL \\
    \hline \hline
    PBR & $ {\{(x_1^\alpha, y_1^\alpha,x_2^\alpha, y_2^\alpha)},(\lambda)\}$ & 96 & 970 &$\{(x_1^\alpha,y_1^\alpha),(x_2^\alpha,y_2^\alpha,\lambda)\}$ & 1\,296 & 4  \\
       &  &  &  & $\{{(x_1^\alpha, y_2^\alpha),(y_1^\alpha,x_2^\alpha,\lambda)}\}$ & 1\,296 &   \\
    PERR & $\{(a_1, a_2), (a_3, a_4),(\lambda)\}$ & 828 &73\,000& $\{(a_1, a_2), (a_3, a_4,\lambda)\}$ & 14\,025 & 6 \\
     &  &  &  &  $\{(a_2, a_3),(a_1,a_3,\lambda)\}$ & 14\,025 &  \\
 
  ICR & $\{(\kappa, \tilde{\kappa}, \tau,\tilde{\tau}) , (\lambda)\}$ &  96  & 61\,870 &$\{(\tilde{\kappa}, \tau), (\kappa, \tilde{\tau}, \lambda)\}$ & 1\,296 & 4 \\
    
	   &  &    &  & $\{(\kappa, \tilde{\tau}), (\tau, \tilde{\kappa}, \lambda)\}$   & 1\,296 &  \\

		\hline
    
     DHR  &  $ \{(e_0, \ldots , t_3)_h,(\lambda,\mu)\}$        &  165\,240  &  -- &  $\{(e_0, \ldots , t_3)_h, (\lambda), (\mu)\}$ & 194\,400 & $\tfrac{11}{10}$ \\
     QBR &    $ \{(e_0, \ldots , t_2)_h,(\lambda,\mu)\}$     &     41\,160 &  -- &  $ \{(e_0, \ldots , t_2)_h,(\lambda), (\mu)\}$  &  82\,320 & $\tfrac{7}{6}$ \\    
    \hline
  \end{tabular}
  \label{tab:6}
\end{table}

To verify the solutions obtained by Bertini, we solved the polynomial system also in Maple 2018 using Gr\"obner basis method. In all cases the Maple solutions matched with those received from Bertini. The total number of solutions for all three representations are summarized in Table~\ref{tab:786}. 
For $\alpha \in$ SE(2)
we get for PERR double the solutions of PBR and ICR as $\pm(a_1,a_2)$  
describe the same rotation. For $\alpha \in$ S(2) we obtain for PERR 86 solutions, where 
76 correspond to the 19 solutions of PBR and ICR as $\pm(a_1,a_2)$ and
$\pm(-a_2i,a_1i)$ result in the same rotation matrix. The remaining 10 solutions 
(given in App.\ A) lie on the 
quadric $a_1^{2}+a_2^{2} = 0$ 
rendering this matrix singular.

\begin{table}[b]
  \centering
  \caption{Total number of solutions (counted including multiplicity) for studied representations}
  \begin{tabular}{|c||c|c|c|c|} \hline
  Motion group &   PBR and ICR & PERR  & BGR and DCKR (Bertini) & BGR and DCKR (Maple)   \\
    \hline\hline
			SE(2)   &   32 & 64  & 162 & \,32 \& 1-dim set \& 2-dim set\,  \\
	 \hline
			S(2)   &   19 & 86   &  & \\
 \hline
  \end{tabular}
  \label{tab:786}
\end{table}

\section{Homogeneous Representations}
\label{HR} 
In this section, we present two homogeneous algebraic representations for each of the motion groups SE(2) and S(2), respectively. 
Two of them are based on Study's kinematic mapping (e.g.\ \cite[p.~86]{husty2013kinematics}), where each element of SE(3) is represented by a point $(e_0:e_1:e_2:e_3:t_0:t_1:t_2:t_3)$ 
in the projective 7-dimensional space $P^7$ located on the so-called Study quadric given by
\begin{equation}
e_0t_0+e_1t_1+e_2t_2+e_3t_3=0
\end{equation}
sliced along the 3-dimensional generator space \medskip $e_0=e_1=e_2=e_3=0$. 

\noindent
\textbf{Blaschke-Gr\"unwald Representation (BGR):} This representation is obtained by restricting Study's parametrization to the planar case {i.e. $e_1=e_2=t_0=t_3=0$. According to \cite[p.~91]{husty2013kinematics} the transformation $\alpha:\,\, \mathbb{R}^2 \to \mathbb{R}^2$
can be written as: 
\begin{equation}\label{eq:bgr}
\alpha:\,\, \mathbf{P}_{i} \mapsto \mathbf{P}_{i}^\alpha:=\frac{1}{\Delta}\left[ \begin{pmatrix}
e_0^2-e_3^2 & -2e_0e_3 \\ 
2e_0e_3 &  e_0^2-e_3^2
\end{pmatrix}\mathbf{P}_{i}+ \mathbf{t}\right] 
\end{equation}
with {$\mathbf{t}:= [-2(e_0t_1-e_3t_2), -2(e_0t_2+e_3t_1)]^T$}, 
$\Delta:=e_0^2+e_3^2$ and $e_0,e_3,t_1,t_2\in\mathbb{R}$. The computation of $e_3$ is based on the Lagrange function $L$ given in \medskip Eq.\ (\ref{s3}).

\noindent
\textbf{Davidson-Hunt Representation (DHR):}  
Based on the analogy of the Study parameters to homogenous screw coordinates (for details see e.g.\ \cite[Section 1.1]{Nawratil2018})
Davidson and Hunt \cite[p.~409]{Davidson2004} suggested to interpret the points in the ambient space $P^7$ of the 
Study quadric as spatial similarity transformations. Following this idea and restricting it to S(2) we end up with 
the following representation of $\alpha:\,\, \mathbb{R}^2 \to \mathbb{R}^2$: 
\begin{equation}
{\alpha}:\,\, \mathbf{P}_{i} \mapsto \mathbf{P}_{i}^\alpha:=\frac{1}{\Delta^2}\left[(\Delta+e_0t_0+e_3t_3) \begin{pmatrix}
e_0^2-e_3^2 & -2e_0e_3 \\ 
2e_0e_3 &  e_0^2-e_3^2
\end{pmatrix} \mathbf{P}_{i}+ \Delta \mathbf{t}  \right]       
\end{equation}
with $e_0,e_3,t_0,t_1,t_2,t_3\in\mathbb{R}$.
The computation of $s_3$ is based on the Lagrange function (\ref{points5b}) with the constraint \medskip
$M:=e_0t_3-e_3t_0$.

\noindent
\textbf{Quaternion Based Representation (QBR):} 
In \cite[Section 2]{Nawratil2016a} a quaternionic formulation of S(4) and S(3) is given, which can also be restricted to S(2). 
Within this approach the transformation $\alpha:\,\, \mathbb{R}^2 \to \mathbb{R}^2$  reads as:
\begin{equation}
{\alpha}:\,\, \mathbf{P}_{i} \mapsto \mathbf{P}_{i}^\alpha:=\frac{1}{\Delta} \left[
 \begin{pmatrix}
e_0f_0-e_3f_3 & -e_0f_3-e_3f_0 \\ 
e_0f_3+e_3f_0 &  e_0f_0-e_3f_3
\end{pmatrix} \mathbf{P}_{i}
+ 
 \mathbf{t}\right]
\end{equation}
with $e_0,e_3,f_0,f_3,t_1,t_2\in\mathbb{R}$.
The computation of $s_3$ is based on the Lagrange function (\ref{points5b}) where the constraint $M$ is given by \medskip
$M:=e_0f_3-e_3f_0$.

\noindent
\textbf{Dual Cayley-Klein Representation (DCKR):} By using the dual Cayley-Klein parameters of SE(2), which are introduced in \cite{nawratil2018parallel}, 
the isotropic point coordinates $(z_i,\overline z_i)$ are  transformed by $\alpha:\,\, \mathbb{C}^2 \to \mathbb{C}^2$ as follows: 
\begin{equation}\label{eq:dckr}
\alpha:\,\, ({z_i},\overline z_i) \mapsto (z_i^\alpha,\tilde z_i^\alpha):=\tfrac{1}{\theta\tilde\theta}\left(
{\theta}(\theta z_{i}+ 2{\sigma}),{\tilde\theta}(\tilde\theta \overline z_{i}+ 2{\tilde\sigma})\right)
\end{equation}
with $ \theta,\sigma, \tilde\theta,\tilde\sigma\in\mathbb{C}$. 
The homogenous 4-tuple $(\theta:\sigma: \tilde\theta:\tilde\sigma)$ corresponds to a real displacement $\alpha$ 
if and only if there exists a value $c\in \mathbb{C}\setminus\left\{0\right\}$ such that $\overline{\theta c}={\tilde\theta c}$ and $\overline{\sigma c}={\tilde\sigma c}$ 
hold, which is equivalent to the condition $\tilde\theta\overline\sigma=\tilde\sigma\overline\theta$.
The computation of $e_3$ is based on the Lagrange function $L$ given in Eq.\ (\ref{s3}).

\begin{remark}
Note that according to \cite{nawratil2018parallel}, BGR and DCKR are linked by the relation:
 $e_0=(\theta+\tilde{\theta})/2$, $e_3 = -i(\theta-\tilde{\theta})/2$, $t_1=(\sigma+\tilde{\sigma})/2$ and $t_2 = -i(\sigma-\tilde{\sigma})/2$. \hfill $\diamond$
\end{remark}

In summary, $\alpha\in$ SE(2) is given by homogenous 4-tuples 
$(e_0:e_3:t_1:t_2)$ in BGR and $(\theta:\sigma: \tilde\theta:\tilde\sigma)$ in DCKR. 
In contrast, $\alpha\in$ S(2) is determined by homogenous 6-tuple 
$(e_0:e_3:t_0:t_1:t_2:t_3)$ in DHR and $(e_0:e_3:f_0:f_3:t_1:t_2)$ in QBR.

\subsection{Results}
\label{RHR}

The following results are again based on the example illustrated in Fig.~\ref{fig1}. 
In all cases the Lagrange function is rational\footnote{The degrees of the polynomials in the numerator and denominator 
with respect to all $n$ unknowns are given as fraction (deg numerator)/(deg denominator) in the column DOL of Tables \ref{tab:5} and \ref{tab:6}.}
 where the polynomials in the numerator and denominator 
are homogenous and of the same degree with respect to to the motion parameters $m_1,\ldots ,m_k$, where 
$k=4$ holds for BGR and DCKR and $k=6$ holds for DHR and QBR. 
Therefore the system of ${n}$ partial derivatives ${L_j}$~$(j=1,\dots, n)$ of ${L}$ results in a homogenous system 
of equations with respect to $m_1,\ldots ,m_k$. 
Due to the homogeneity this polynomial system is overdetermined but it can easily be checked that the relation 
$\sum_{j=1}^k m_j\frac{\partial L}{\partial m_j}=0$ 
holds. Bertini cannot handle overdetermined systems\footnote{In contrast to the software package HomotopyContinuation.jl~\cite{10.1007/978-3-319-96418-8_54}.} without the following user interaction:  
One has to square up the system \cite[p.~14]{bates2013numerically} by replacing the $k$ equations $\frac{\partial L}{\partial m_j}=0$ for $j=1,\ldots ,k$ by 
$k-1$ linear combinations of the form
$\sum_{j=1}^k \square\frac{\partial L}{\partial m_j}$ 
where each $\square$ indicates a randomly chosen complex number. The resulting system can then be passed on to Bertini by using the command 
{\it hom\_variable\_group} for grouping the homogenous variables $m_1,\ldots ,m_k$ into one group, which is indicated in 
Tables \ref{tab:5} and \ref{tab:6} by the notation $(m_1,\ldots ,m_k)_h$. 
Therefore there is only one possible grouping for BGR and DCKR and two possible groupings for DHR and QBR. 
For both of these groupings the resulting B\'{e}zout numbers are too large (cf.\ Table \ref{tab:6}) 
to expect reasonable computation times thus we abstained from tracking the paths.   

All additional 130 solutions received by Bertini based on BGR and DCKR (cf.\ Table  \ref{tab:786}) 
fulfill $\Delta=0$ resp.\ $\theta\tilde\theta=0$ implying a 
division by zero in Eq.\  (\ref{eq:bgr}) resp.\ (\ref{eq:dckr}). But it turns out that 128 of these solutions 
either result from squaring up the system \cite[p.~15]{bates2013numerically} (cf.\ Apps.\ B and C) or
belong to the 2-dimensional solution set\footnote{Positive-dimensional solution sets cannot be detected by Bertini using default settings.\label{reml}}  $e_0=e_3=0$ resp.\ $\theta=\tilde\theta=0$. 
There also exists a 1-dimensional solution set \footnotemark[\getrefnumber{reml}] (cf.\ Apps.\ B and C), 
which corresponds to two conics on the 
hyperquadric $\Delta=0$ resp.\ $\theta\tilde\theta=0$. The remaining two solutions represent a point on each of these two curves.

\section{Conclusions}
\label{conc} 
It can be observed (cf.\ Table \ref{tab:5}) that the $B_{min}$ value obtained for ICR is the best one of all SE(2) representations as suggested in \cite [p.~96]{bates2013numerically}. For S(2) the lowest number of tracked paths is obtained by ICR and PBR (cf.\ Table \ref{tab:6}). 
Surprisingly the PBR has in both motion groups by far the best computational performance with respect to $T_{avg}$. 
It can also be seen by $B_{max}$ that the grouping has huge effects on the number of tracked paths and therefore on the computation time. 
Moreover, due to the large  B\'{e}zout numbers of DHR and QBR the question arises whether a computationally more efficient homogenous representation of S(2) exists. 

All in all, this study suggests the usage of PBR for the future research on the spatial case 
(i.e.\  hexapods and linear pentapods)  
due to the good $B_{min}$ values and the best results for $T_{avg}$.

\begin{acknowledgement}
This research is supported by the Grant No.P 30855-N32 of the Austrian Science Fund FWF. The  first author would like to thank Arvin Rasoulzadeh for his valuable suggestions and technical discussions.	
\end{acknowledgement}

\bibliographystyle{spmpsci}
\bibliography{bibfile1}

\section*{Appendix}

\subsection*{A. Additional ten solutions of PERR under equiform motions}

In the following we give the explicit expressions of the 10 solutions mentioned in Sec.\ \ref{RNHR} 
lying on the quadric \smallskip $a_1^{2}+a_2^{2} = 0$.

\begin{figure}[t]
\begin{center}
\begin{overpic}[width = 85mm]{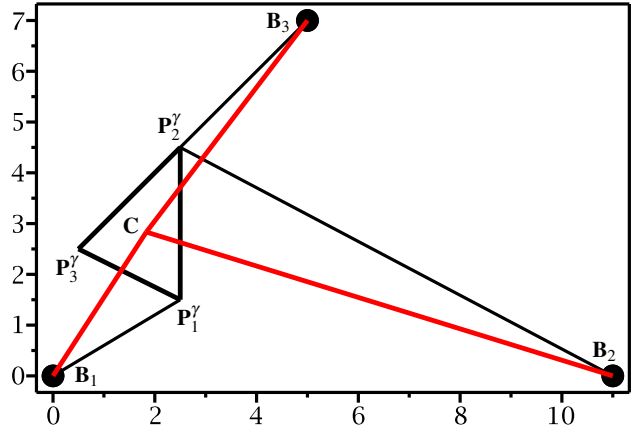}
\put(12,21){$\mathbf{B}_{1}$}‎
\put(28,30){$\mathbf{P}^\gamma_{1}$}‎
\put(93,25){$\mathbf{B}_{2}$}‎
\put(25,60){$\mathbf{P}^\gamma_{2}$}‎
\put(9,37.8){$\mathbf{P}^\gamma_{3}$}‎
\put(19.5,44.5){$\mathbf{C}$}‎
\put(42,76.5){$\mathbf{B}_{3}$}‎
\end{overpic}
\vspace{-3em}
\caption{Illustration of the double solution: 
The platform degenerates to the centroid $\mathbf{C}$ (yielding the red singular configuration) of the moving platform in the given pose (displayed in black).}
\label{fig2}
\end{center}
\end{figure}

\noindent
\textbf{Solutions 1 and 2:}  The real solution $a_1=0$, $a_2=0$, $a_3=\frac{11}{6}$, $a_4=\frac{17}{6}$  is of multiplicity 2. 
In this case the platform degenerates to the centroid $\mathbf{C}$ of the moving platform in the given pose \smallskip (cf.\ Fig.~\ref{fig2}).

\noindent
\textbf{Solutions 3--6:} The third solution is given by: 
\begin{align*}
&a_1=\tfrac{\sqrt{w_0}\sqrt{(w_1-w_2i)+\sqrt{\tfrac{(w_3+w_4i)}{w_5}}}}{w_6}, &\quad 
&a_3=u_{0}+u_{1}i-\tfrac{b_1\sqrt{w_3+w_4i}}{k_1}-\tfrac{n_{1}i\sqrt{-w_3-w_4i}}{k_2},\\
&a_2= \tfrac{-i\sqrt{(r_0-r_1i)+2\sqrt{\tfrac{(w_3+w_4i)}{w_5}}}}{w_6}, &\quad 
&a_4=q_{0}-q_{1}i-\tfrac{b_2\sqrt{w_3+w_4i}}{k_3}+\tfrac{n_{2}i\sqrt{-w_3-w_4i}}{k_4}
\end{align*}
with $w_0 =2$,  $w_1 =31779207$, $w_2=30431636$, $w_3=120733747144604$, $w_4=831183587338$, $w_5=21040857161$, $w_6=3796$, $r_0=63558414$,  $r_1=60863272$, $u_0=-\tfrac{17446123}{3602404}$, $u_1=\tfrac{46887293}{1801202}$, $q_0=-\tfrac{21374427}{3602404}$, $q_1=\tfrac{1811805}{900601}$  $b_1=18854507559$, $b_2=52996212054$, $n_1=52996212054117$ and $n_2=14299077184932$.
The other three solutions for $(a_1,a_2,a_3,a_4)$ equal $(-a_1,-a_2,a_3,a_4)$ and $(\pm a_2i,\mp a_1i,a_3,a_4)$, respectively. 
\smallskip 

\noindent
\textbf{Solutions 7--10:} The seventh solution is given by: 
\begin{align*}
&a_1=\tfrac{\sqrt{w_0}\sqrt{(w_1+w_2i)+\sqrt{\tfrac{(-w_3+w_4i)}{w_5}}}}{w_6}, &\quad
&a_3=u_{0}-u_{1}i-\tfrac{b_1\sqrt{-w_3+w_4i}}{g_1}-\tfrac{n_{1}i\sqrt{-w_3+w_4i}}{g_2},\\
&a_2= \tfrac{i\sqrt{(r_0+r_1i)+2\sqrt{\tfrac{(-w_3+w_4i)}{w_5}}}}{w_6}, &\quad
&a_4=q_{0}+q_{1}i-\tfrac{b_2\sqrt{-w_3+w_4i}}{g_3}+\tfrac{n_{2}i\sqrt{-w_3+w_4i}}{g_4}
\end{align*}
with
$g_1=772036659995716$, $g_2=38601832999$, $g_3=77203665999$ and $g_4=19300916499$. 
The other three solutions for $(a_1,a_2,a_3,a_4)$ equal $(-a_1,-a_2,a_3,a_4)$ and $(\pm a_2i,\mp a_1i,a_3,a_4)$, respectively.

\subsection*{B. Computational details of BGR}

\noindent
\textbf{Maple computations:}
It can be verified using Gr\"obner basis method that the polynomial system does not only have 32 isolated solutions and the 2-dimensional solution set $e_0=e_3=0$ but also a 1-dimensional solution set given by:
\begin{equation}\label{conic:bgr}
\begin{split}
&e_0=1,\quad e_3=\pm i \quad\text{and} \\
&3752e_3t_1-13723e_3t_2+1865t_{1}^{2}+1865t_{2}^2+8629t_{1}+2144t_{2}=0.
\end{split}
\end{equation}
Therefore it corresponds to two conic sections on the quadric \smallskip  $e_0^2+e_3^2=0$.

\noindent
\textbf{Bertini computations:}
We squared up the system of equations as described in Sec.\ \ref{RHR} by using the numerators of the following three linear combinations: 
\begin{align*}
&(1-7i)\tfrac{\partial L}{\partial e_0}+(2-3i)\tfrac{\partial L}{\partial e_3}+(3+5i)\tfrac{\partial L}{\partial t_1}+(4-6i)\tfrac{\partial L}{\partial t_2},\\
&(4+2i)\tfrac{\partial L}{\partial e_0}+(3-3i)\tfrac{\partial L}{\partial e_3}+(3-2i)\tfrac{\partial L}{\partial t_1}+(3+5i)\tfrac{\partial L}{\partial t_2},\\
&(2+3i)\tfrac{\partial L}{\partial e_0}+(3-5i)\tfrac{\partial L}{\partial e_3}+(2-2i)\tfrac{\partial L}{\partial t_1}+(3-3i)\tfrac{\partial L}{\partial t_2}.
\end{align*}
Based on this input the 130 additional solutions obtained by Bertini split up in the following way: We get one solution 
on each of the conic sections given in Eq.\ (\ref{conic:bgr}), 39 solutions belong to the 2-dimensional solution set $e_0=e_3=0$ and 89 solutions 
result from squaring up the system (they do not fulfill the initial set of equations). 

\begin{remark}\label{rem:seed}
All 130 additional solutions depend on the random seed chosen by Bertini; even the number of solutions belonging to the 
2-dimensional set is not constant. But it turns out in all our examples that exactly two solutions are located on the 
1-dimensional set (one on each conic section).  \hfill $\diamond$
\end{remark}

\subsection*{C. Computational details of DCKR}

\noindent
\textbf{Maple computations:}
The polynomial system has 32 isolated solutions, the 2-dimensional solution set
$\theta=\tilde\theta=0$ and a 1-dimensional solution set, which corresponds to the following two conic sections:
\begin{equation} \label{conic:dckr}
\begin{split}
&\theta=0, \quad \tilde{\theta}=1, \quad
1865\tilde{\sigma}\sigma+2547\tilde{\sigma}+804i\tilde{\sigma}-11176\sigma+2948i\sigma=0, \\
&\theta=1, \quad \tilde{\theta}=0, \quad
1865\tilde{\sigma}\sigma+2547\sigma-804i\sigma-11176\tilde{\sigma}-2948i\tilde{\sigma}=0.
\end{split}
\end{equation}

\noindent
\textbf{Bertini computations:}
We squared up the system of equations as described in Sec.\ \ref{RHR} by using the numerators of the following three linear combinations: 
\begin{align*}
&(3-i)\tfrac{\partial L}{\partial \theta}+(3-4i)\tfrac{\partial L}{\partial \tilde{\theta}}+(2-3i)\tfrac{\partial L}{\partial \sigma}+(5-4i)\tfrac{\partial L}{\partial \tilde{\sigma}},\\
&(1-2i)\tfrac{\partial L}{\partial \theta}+(3-5i)\tfrac{\partial L}{\partial \tilde{\theta}}+(2-3i)\tfrac{\partial L}{\partial \sigma}+(5-5i)\tfrac{\partial L}{\partial \tilde{\sigma}},\\
&(7-i)\tfrac{\partial L}{\partial \theta}+(3-2i)\tfrac{\partial L}{\partial \tilde{\theta}}+(1-2i)\tfrac{\partial L}{\partial \sigma}+(5-2i)\tfrac{\partial L}{\partial \tilde{\sigma}}.
\end{align*}
Based on this input the 130 additional solutions obtained by Bertini split up in the following way: We get one solution 
on each of the conic sections given in Eq.\ (\ref{conic:dckr}), 41 solutions belong to the 2-dimensional solution set $\theta=\tilde\theta=0$ and 87 solutions 
result from squaring up the system (they do not fulfill the initial set of equations). 

Finally it should be mentioned that Remark \ref{rem:seed} is also valid for this case.

\end{document}